\documentclass[aps,prl,twocolumn,showpacs,floatfix,superscriptaddress]{revtex4}
\usepackage{graphicx}
\usepackage{hyperref}
\usepackage{amsmath}

\usepackage{color}

\newcommand{\p}[1]{#1^\dagger} 
\newcommand{\expo}[1]{e^{#1}} 
\newcommand{\ket}[1]{|{#1}\rangle}

\begin{document}
\flushbottom
\title{Emitters of $N$-photon bundles}

\author{C. S\'anchez Mu\~noz}
\affiliation{Condensed Matter Physics Center (IFIMAC), Departamento de F\'isica Te\'orica de la Materia Condensada, Universidad Aut\'onoma de Madrid, 28049 Madrid, Spain}

\author{E. del Valle}
\affiliation{Condensed Matter Physics Center (IFIMAC), Departamento de F\'isica Te\'orica de la Materia Condensada, Universidad Aut\'onoma de Madrid, 28049 Madrid, Spain}
\affiliation{Physik Department, Technische Universit\"at M\"unchen, James Franck Str., 85748 Garching, Germany}

\author{A. Gonz\'alez Tudela}
\affiliation{Max-Planck-Institut f\"ur Quantenoptik, Hans-Kopfermann-Str.~1, 85748 Garching, Germany}

\author{S.~Lichtmannecker}
\affiliation{Walter Schottky Institut, Technische Universit\"at M\"unchen, Am Coulombwall 4, 85748 Garching, Germany}

\author{K. M\"uller}
\affiliation{Walter Schottky Institut, Technische Universit\"at M\"unchen, Am Coulombwall 4, 85748 Garching, Germany}

\author{M. Kaniber}
\affiliation{Walter Schottky Institut, Technische Universit\"at M\"unchen, Am Coulombwall 4, 85748 Garching, Germany}

\author{C. Tejedor}
\affiliation{Condensed Matter Physics Center (IFIMAC), Departamento de F\'isica Te\'orica de la Materia Condensada, Universidad Aut\'onoma de Madrid, 28049 Madrid, Spain}

\author{J.J. Finley}
\affiliation{Walter Schottky Institut, Technische Universit\"at M\"unchen, Am Coulombwall 4, 85748 Garching, Germany}

\author{F.P. Laussy}
\affiliation{Condensed Matter Physics Center (IFIMAC), Departamento de F\'isica Te\'orica de la Materia Condensada, Universidad Aut\'onoma de Madrid, 28049 Madrid, Spain}
\email{fabrice.laussy@gmail.com}

\begin{abstract}
  We propose a scheme based on the coherent excitation of a two-level
  system in a cavity to generate an ultrabright CW and focused source
  of quantum light that comes in groups (bundles) of $N$~photons, for
  an integer~$N$ tunable with the frequency of the exciting laser. We
  define a new quantity, the \emph{purity} of $N$-photon emission, to
  describe the percentage of photons emitted in bundles, thus
  bypassing the limitations of Glauber correlation functions. We focus
  on the case $1\le N\le3$ and show that close to 100\% of two-photon
  emission and 90\% of three-photon emission is within reach of state
  of the art cavity QED samples.  The statistics of the bundles
  emission shows that various regimes---from $N$-photon lasing to
  $N$-photon guns---can be realized.  This is evidenced through
  generalized correlation functions that extend the standard
  definitions to the multi-photon level.
\end{abstract}
 \date{\today} \maketitle

Cavity Quantum Electrodynamics (cQED) allows to control the
interaction of light with matter at the ultimate quantum
limit~\cite{haroche_book06a} with prospects for engineering new
generations of light sources~\cite{walther06a,obrien09a}.  In this
Letter, we propose a family of \emph{$N\!$-photon emitters}, i.e.,
sources that release their energy exclusively in bundles of
$N$~photons (for integer $N$). The statistics of emission of the
bundles can be varied with system parameters from antibunching to
poissonian, thereby realizing $N$-photon guns and $N$-photon lasers.
Such non-classical emitters are highly sought for robust quantum
information processing, generating NOON states~\cite{afek10a}, quantum
lithography and metrology~\cite{giovannetti04a}, but also for medical
applications, allowing for higher penetration lengths and increased
resolution with minimum harm to the
tissues~\cite{denk90a,horton13a}. The recent demonstration that
biological photoreceptors are sensitive to photon
statistics~\cite{sim12a} may also render such sources highly relevant
for studies of biological photosystems and, potentially, of quantum
biology.

\begin{figure}[t]
  \includegraphics[width=.8\linewidth]{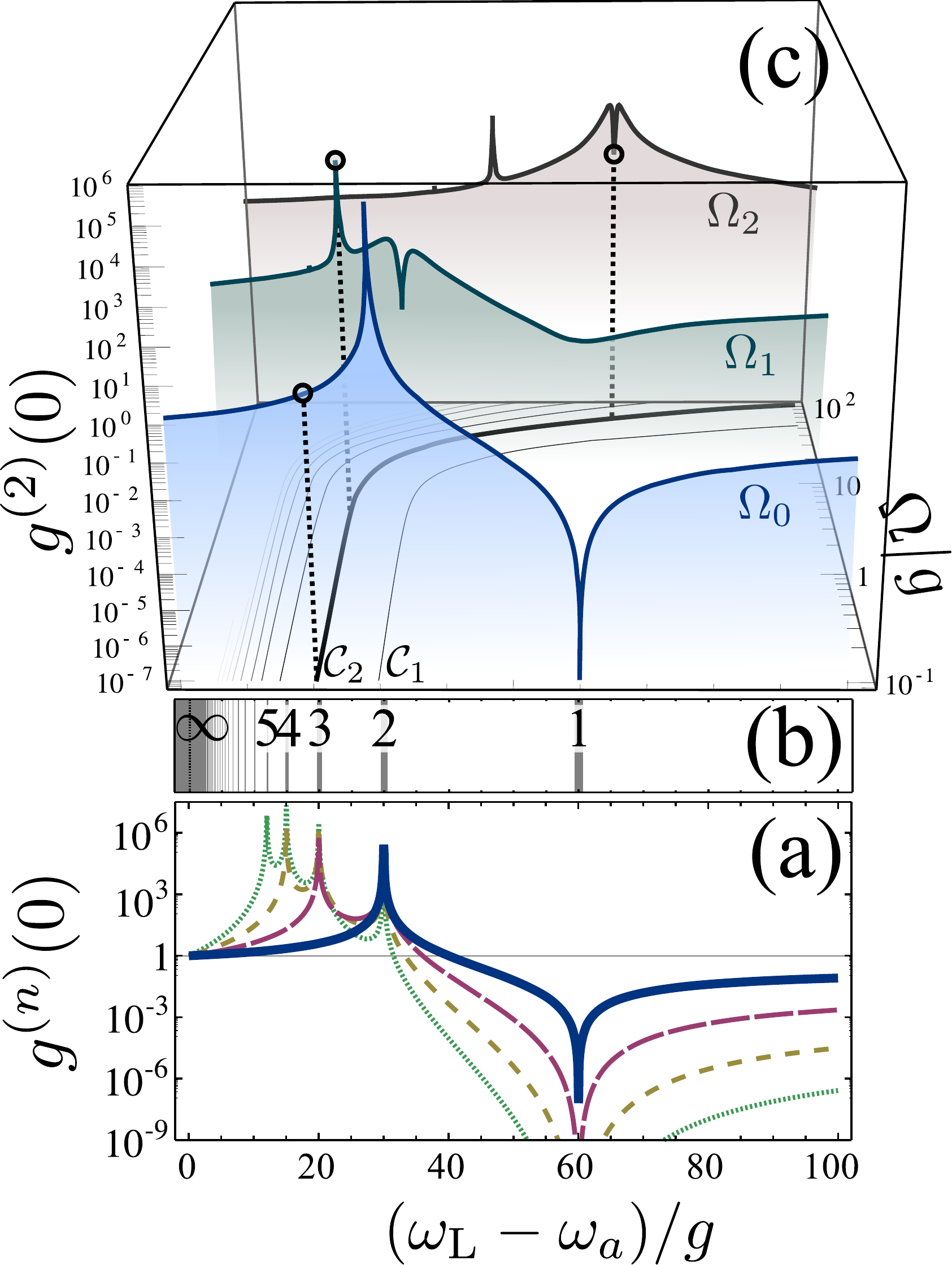}
  \caption{(Colour online) From bottom to top: (a) $g^{(n)}$ for $n=2$ (solid), 3 (long dash), 4 (short dash) and 5 (dotted) at vanishing pumping with $n-1$ bunching resonances matching
    those in (b), the resonant energies to excite the $n$th rung of
    the ladder when $\Delta/g=-60$. (c) $g^{(2)}$ as a function of
    $\omega_\mathrm{L}$ for pumping $\Omega_0\approx10^{-2}g$,
    cf.~(a), $\Omega_1\approx4g$ and $\Omega_2\approx32g$. The
    resonances $\mathcal{C}_N$ are shown in the plane
    $(\omega_\mathrm{L},\Omega)$. Open circles are the projection of
    $\mathcal{C}_2$ on $g^{(2)}$.}
  \label{fig:2}
\end{figure}

Our scheme relies on the paradigm of cQED: one two-level system in a
cavity. This is realized in a wealth of physical systems, ranging from
atoms in optical cavities~\cite{brune96a} to superconducting qubits in
microwave resonators~\cite{fink08a} and quantum dots in
microcavities~\cite{kasprzak10a}.  The dynamics is well described by
the Jaynes--Cummings Hamiltonian $H_0 = \omega_a\p{a}a +
\omega_\sigma\p{\sigma}\sigma + g(\p{a}\sigma + \p{\sigma}a)$ with $a$
and $\sigma$ the second quantization annihilation operators of the
light field (boson statistics) and the Quantum Emitter (QE, two-level
system), respectively, with corresponding free energies $\omega_a$ and
$\omega_\sigma$ and coupling strength $g$~\cite{jaynes63a}.
\begin{figure}[th!]
  \includegraphics[width=.8\linewidth]{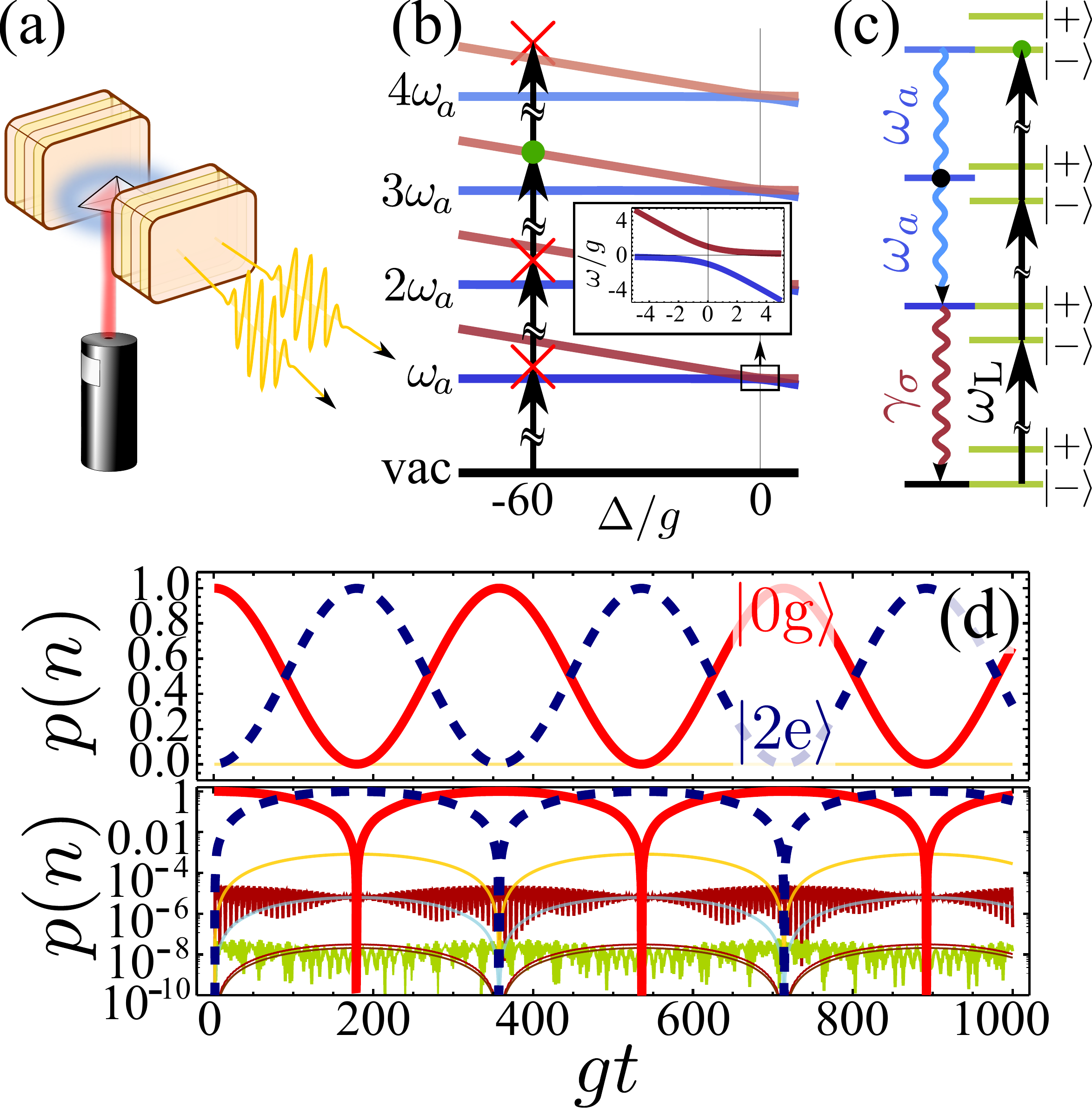}
  \caption{(Colour online) (a) Scheme of our system: coherent
    excitation of a two-level system far detuned in energy from the
    mode of a cavity emitting $N$-photon bundles. (b) Energy scheme in
    the low excitation regime (anticrossing magnified in inset)
    resonantly exciting the third rung of the ladder with photon
    blockade at all others. (c) Energy scheme in the high excitation
    regime: the laser dresses the QE while the cavity Purcell-enhances
    a two-photon transition from $\ket{-}$ to $\ket{+}$. (d) Full
    amplitude Rabi oscillations realised in the conditions of
    panel~(b), in linear and log scales. }
  \label{fig:1}
\end{figure}
The configuration under study is the resonant excitation by an
external laser of the QE~\cite{nguyen11a,matthiesen12a,jayakumar13a}
far in the dispersive regime with the cavity
($|\omega_a-\omega_\sigma| \gg
g\sqrt{N+1}$)~\cite{majumdar10a,hughes11a,laussy12e}
(Fig.~\ref{fig:1}(a)).  The energy structure of~$H_0$ is shown in
Fig.~\ref{fig:1}(b) with the QE at $\Delta/g=-60$
($\Delta=\omega_a-\omega_\sigma$), in which case the states are
essentially the bare ones. The laser of frequency~$\omega_\mathrm{L}$
and pumping intensity $\Omega$ is included by adding
$\Omega(\expo{-i\omega_\mathrm{L}t}\p{\sigma}+\expo{i\omega_\mathrm{L}
  t}\sigma)$ to $H_0$. At pumping low enough not to distort the level
structure, one can excite selectively a state with $N$ photon(s) in
the cavity at the $(N+1)$th rung by adjusting the laser frequency to
satisfy
\begin{equation}
  \label{eq:juesep5202411CEST2013}
  \omega_\mathrm{L}=\omega_a+\frac{\sqrt{4(N+1)g^2+\Delta^2}-\Delta}{2(N+1)}\,,
\end{equation}
with~$N\in\mathbf{N}$~\cite{chough00a}. This is shown in the figure~\ref{fig:1}(b) for the case $N=2$,
with a photon-blockade~\cite{birnbaum05a,faraon08a} at all other rungs
(above and below)~\cite{schuster08a,bishop09a}. The positions of the
resonances in the case $\Delta/g=-60$ are shown in Fig.~\ref{fig:2}(b).
Rightmost is the resonant excitation of the QE while other resonances
pile up towards the cavity frequency, showing a transition from the
quantized features of the QE towards the classical continuum of the
cavity. In the absence of dissipation, this leads to the generation of
an exotic brand of maximally entangled polaritons, of the type
$(\ket{0\mathrm{g}}\pm\ket{N\mathrm{e}})/\sqrt{2}$ rather than the
usual case $(\ket{0\mathrm{e}}+\ket{1\mathrm{g}})/\sqrt{2}$. The
dynamics of the system driven upon exciting the case $N=2$,
corresponding to the excitation of the third rung, is presented in
Fig.~\ref{fig:1}(c), both in linear and log-scales.  Full amplitude
Rabi oscilations between the $\ket{0\mathrm{g}}$ and
$\ket{N\mathrm{e}}$ states are observed. 

When increasing pumping, resonances in the amplitude of the Rabi
oscillations persist but are blueshifted due to the dressing of the
Jaynes--Cummings states by the laser. The level structure becomes that
of a dressed atom~\cite{cohentannoudji77a} strongly detuned from a
cavity mode~\cite{zakrzewski91a}, leading to resonances at:
\begin{equation}
  \label{eq:juesep5201806CEST2013}
  \omega_\mathrm{L}=\omega_a+\frac{\sqrt{4(N^2-1)\Omega^2+N^2\Delta^2}+\Delta}{N^2-1}\,,
\end{equation}
realized when the energy of $N$ cavity-photons match the $N$-photon
transition between the $\ket{-}$ and $\ket{+}$ levels of the dressed
atom, as sketched in Fig.~\ref{fig:1}(c) for the case $N=2$. In the
indeterminate case $N=1$, Eq.~\eqref{eq:juesep5201806CEST2013} should
be taken in the limit $N\rightarrow 1$, yielding $\omega_L =
\omega_a-(2 \Omega^2+ \Delta^2/2)/\Delta$ (in the dispersive regime,
$\Delta \neq 0$).  All the dynamics discussed so far correspond to
systems that are Hamiltonian in nature, such as atomic cQED
realizations~\cite{gleyzes07a}.

\begin{figure*}[t]
  \includegraphics[width=\linewidth]{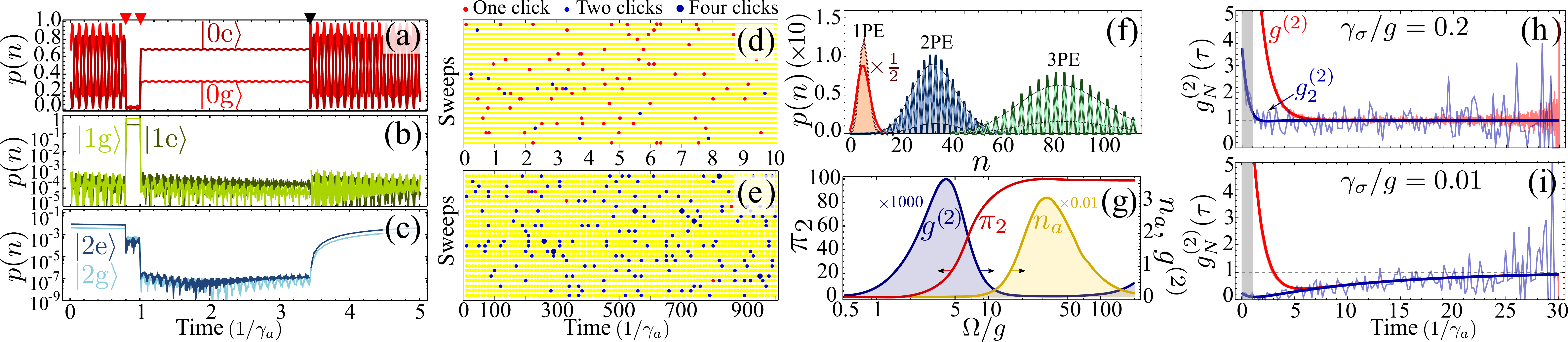}
  \caption{(a--c) Quantum trajectory showing the probability of the
    system to be in any of the states $\ket{n\mathrm{g/e}}$ when
    exciting the third rung.  (d--e) Cavity-photon clicks as they
    would be recorded by a streak camera (25 sweeps shown) for the
    pumping values $\Omega_1$ (d) and $\Omega_2$ (e) at
    $\mathcal{C}_2$. In (d) the emission is highly bunched although it
    largely consists of single clicks, $g^{(2)}=3649$ and
    $\pi_2=16\%$, while in (e), $g^{(2)}=17$ with $\pi_2=98.8\%$. (f)
    Ideal $N$PE ($N$-Photon Emission) in thick lines and 99\% $N$PE in
    translucid lines with an envelope to guide the eye. (g) Pumping
    dependence of, left axis, $\pi_2$ and, right axis, $g^{(2)}$ (from
    0 to $3\,649$) and $n_a$ (from 0 to $0.03$) following
    $\mathcal{C}_2$. (h--i) Photon correlations at the $N=1$ (red)
    and~$N=2$ (blue) level, from Eq.~(\ref{eq:mieoct9203726CEST2013})
    (smooth curve) and from Monte Carlo clicks (data). Small
    deviations from the ideal case occur in the small time window
    $1/\gamma_a$. Antibunching and coherent emission of photon pairs
    are otherwise realized.}
  \label{fig:3}
\end{figure*}

Strong dissipation, e.g., in semiconductor cQED, is not detrimental to
quantum effects~\cite{volz12a,majumdar12a}.  On the contrary, Purcell
enhancement of such Hamiltonian resonances may give rise to giant
photon correlations in the statistics of the field detected
\emph{outside} the cavity instead of Rabi
oscillations~\cite{delvalle11d,zubairy80a}. The corresponding
zero-delay photon correlations $g^{(n)}$~\cite{glauber63b} are shown
in the limit of vanishing pumping in Fig.~\ref{fig:2}(a). An
antibunching dip is observed for each $g^{(n)}$ when exciting
resonantly the emitter, followed by a series of $N-1$ huge bunching
peaks that match precisely the resonances
Eq.~(\ref{eq:juesep5202411CEST2013}), plotted in
Fig.~\ref{fig:2}(b). In these calculations, as for all the following
results, the Hamiltonian has been supplemented with superoperators in
the Lindblad form to describe dissipation of the cavity (resp.~QE) at
a rate $\gamma_a$ (resp.~$\gamma_\sigma$)~\cite{kavokin_book11a}.
Parameters used for the examples shown here are $\gamma_a/g=0.1$
and~$\gamma_\sigma/g=0.01$. As pumping is increased, resonances in
$g^{(n)}$ shift, as expected, along curves $\mathcal{C}_N$ in the
$(\omega_\mathrm{L},\Omega)$ space defined by
Eq.~(\ref{eq:juesep5201806CEST2013}). This is shown for $g^{(2)}$ in
Fig.~\ref{fig:2}(c) for three values of pumping, starting with
$\Omega_0=10^{-1}g$, close to the vanishing pumping case shown in
Fig.~\ref{fig:2}(b). Following $g^{(2)}$ along the $\mathcal{C}_2$
resonance shows that a new peak emerges out of a uniform background,
reaching a maximum $g^{(2)}\approx3649$ at the pumping
$\Omega_1\approx4g$ (middle trace) before a depletion of the resonance
forms for higher pumping, reaching its minimum along $\mathcal{C}_2$
of $g^{(2)}\approx 17$ at $\Omega_2\approx32g$ (background trace).

Such resonances are indicative of strong correlations but not in an
intuitive way nor in a particularly useful one for applications, since
$g^{(2)}$ is unbounded and cannot be interpreted in terms of
probability of two-photon emission. Other quantities to measure
correlations, such as the differential correlation
function~\cite{kubanek08a} or the surge~\cite{hong10a}, present the
same problem. To gain insights into the dissipative context, we turn
to a quantum Monte Carlo approach~\cite{plenio98a}, where one can
follow individual trajectories of the system and record photon clicks
whenever the system undergoes a quantum jump. A tiny fraction of such
a trajectory when the third rung is excited is presented in
Figs.~\ref{fig:3}(a--c), showing the probabilities of the system to be
in the states $\ket{n\mathrm{g}/\mathrm{e}}$ for $n$ up to~2
(probabilities in higher rungs are included in the numerical
simulation but do not play a role in the dynamics).  Until time
$t\approx0.9$ (in units of $1/\gamma_a$), the QE essentially undergoes
fast Rabi flopping (in an empty cavity) under the action of the laser,
corresponding to the Mollow regime. At the same time, the driving of
the third rung makes the probability to have two photons in the cavity
sizable, as can be seen in Fig.~\ref{fig:3}(c) where the combined
probability reaches over 1\%, while the probability to have one photon
is more than two orders of magnitude smaller.  The high probability of
the two-photon state eventually results in the emission of a first
cavity photon indicated by a red triangle at the top of the figure,
that collapses the wavefunction into the one-photon state, that is now
the state with almost unit probability. While only the continued Rabi
flopping of the QE was extremely likely at any moment of time before
the first photon emission, the system is now expected to emit a second
photon within the cavity lifetime of the first one. This corresponds
to the second red triangle in Fig.~\ref{fig:2}(a) denoting the second
photon emission within $0.2/\gamma_a$ of its precursor. There is a
jitter in the emission of the two-photon state due to the cavity, but
this does not destroy their correlation. Phrased equivalently, the
two-photon picture holds outside of a small time window constrained by
the cavity decay rate.  After the two-photon emission, the system is
left in a vacuum state but without Rabi flopping, that is restored
after a direct emission from the QE (black triangle) and a two-photon
state is again constructed, preparing for the next emission of a
correlated photon pair.  The system is brought back to its starting
point, conserving total energy over a cycle, by direct emission from
the QE. Although one QE photon is emitted per two-photon emission
cycle, it is at another frequency and in a different solid angle. The
two-photon emission is through the cavity mode, being therefore
strongly focused.

It must be stressed that such quantum dynamics producing cavity
photons in pairs does not correspond to the huge bunching resonances
in $g^{(2)}$ in Fig.~\ref{fig:2}(c). Here, the system indeed emits
more frequently two photons as compared to a random source with the
same intensity, but with an overwhelming predominance of single
photons, as is observed when considering the actual emission of the
system. Figures~\ref{fig:3}(d--e) present a series of detection events
such as they would be recorded by a streak camera
photodetector~\cite{wiersig09a}, calculated by the quantum Monte Carlo
method for the pumping values $\Omega_1$ and $\Omega_2$ of
Fig.~\ref{fig:2}(c) at $\mathcal{C}_2$. The horizontal axis represents
time and each row of points denotes a detection event as the detection
spot is raster scanned accross the image.  The red points correspond
to one photon detection events and blue points to correlated
two-photon emissions.  Indeed, while the strong bunching in
Fig.~\ref{fig:2}(c) at $\Omega_1$ only conveys that the number of
correlated two-photon events (blue points) in Fig.~\ref{fig:3}(d) is
much larger than would be expected for a coherent source, the
emission remains predominantly in terms of single photons.  Whilst the
resonances in statistics are strong, they are therefore not meaningful
for applications. On the other hand, at $\Omega_2$, when the $g^{(2)}$
resonance is depleted, the emission now consists almost exclusively of
correlated photon pairs, as can be seen by the dominance of blue
points in Fig.~\ref{fig:3}(e).
\begin{figure}[t]
  \includegraphics[width=.9\linewidth]{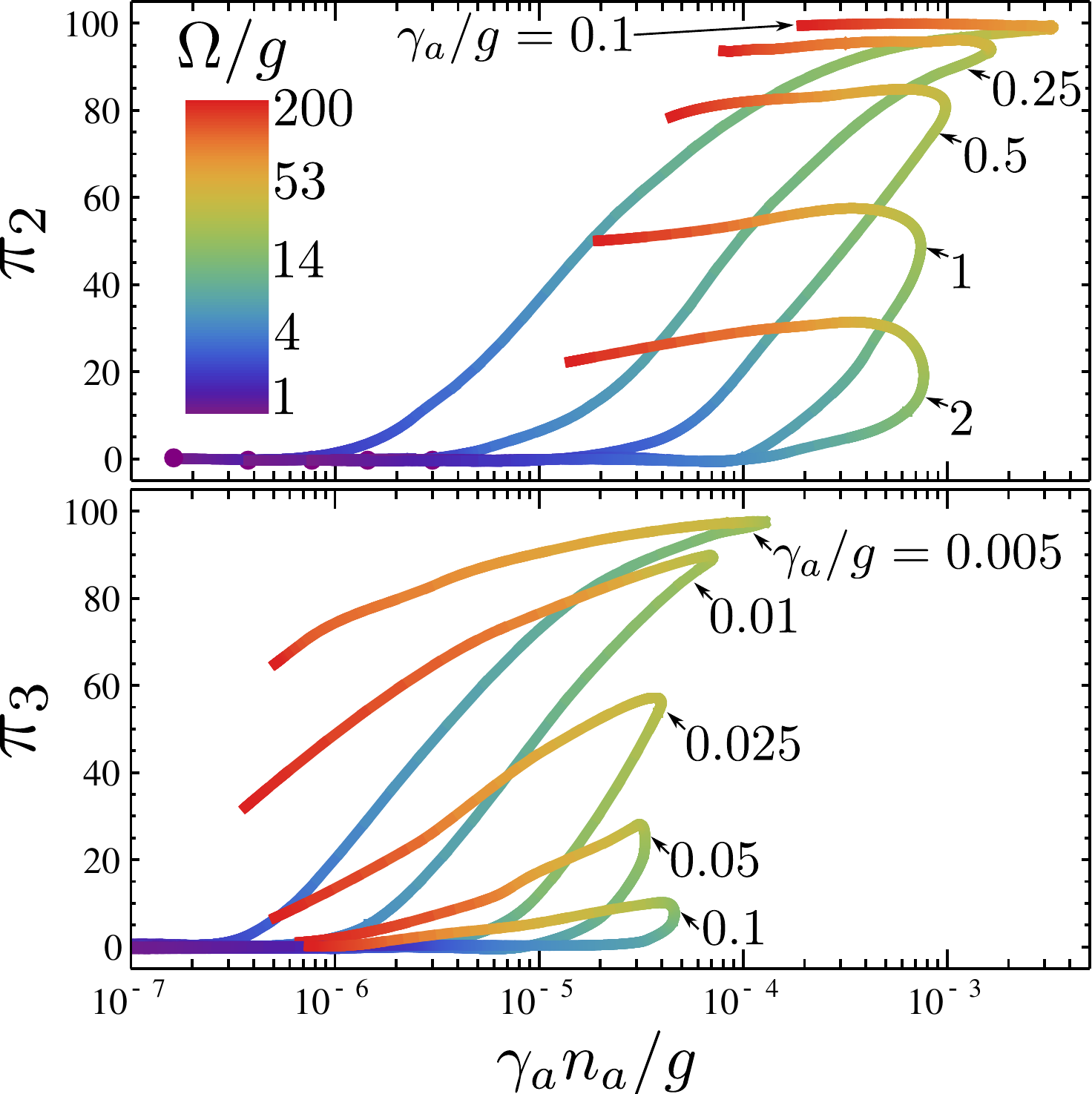}
  \caption{Figures of merit for two- and three-photon emissions in the
    space of purity/emission intensity. Almost pure two-photon and
    three-photon emission can be achieved with state of the art cQED
    samples: $\gamma_\sigma/g=0.01$ for~$\pi_2$ and $0.001$
    for~$\pi_3$.}
  \label{fig:4}
\end{figure}
The standard correlation functions~$g^{(n)}$ are therefore not
suitable to describe the physics of $N$-photon
emission. Photon-counting~\cite{srinivas81a,zoller86a,osadko09a} is a
convenient way to quantify the amount of $N$-photon emission in
practical terms, since an ideal $N$-photon emitter never produces a
number of photons which is not a multiple of
$N$~\cite{carmichael93a}. We observe that for time windows $T$ much
larger than the coherence time, counting of the photon bundles becomes
poisson distributed, as short time correlations are
lost~\cite{loudon_book00a}. In that limit, the random variable $X_N$
that counts bundles of $N$ photons in a time window~$T$ is
$P(X_N=k)=\exp(-\lambda_N T)(\lambda_N T)^{k/N}/(k/N)!$ if $k$ is a
multiple of $N$, and is zero otherwise, with a generating function
$\Pi_{X_N}(s)=\langle s^{X_N}\rangle=e^{-\lambda_N(1-s^N)}$. This
distribution is shown in Fig.~{\ref{fig:3}(f) for the cases of ideal
  two-photon (2PE) and three-photon (3PE) emission. The latter is
  shown for a larger counting time window to shift the distribution
  sidewise. A non-ideal $N$-photon emitter occasionally emit single
  photons that spoil these distributions. Photon counting then results
  from the sum of two random variables $X_1+X_N$ where $X_1$ is a
  conventional Poisson process. The generating function of the
  imperfect $N$-photon emitter is $\Pi_{X_1+X_2}=\Pi_{X_1}\Pi_{X_2}$
  with distribution to count $n$ photons in the time window~$T$:
\begin{equation}
  \label{eq:TueMay14112405CEST2013}
  P_N(n)=e^{-(\lambda_1+\lambda_N)T}\sum_{k=0}^{n}\frac{n!(\lambda_1 T)^{n-Nk}(\lambda_N T)^k}{k!(n-Nk)!}\,.
\end{equation}
When the suppression of photon emission that is not a multiple of $N$
is efficient, we find these parameters to be related to the cavity
population through $\lambda_N=\gamma_an_a/N$.  The $\lambda$
parameters being independent of the time window~$T$, are suited
to characterize $N$-photon emission.  Therefore, we define the
\emph{purity} of $N$-photon emission, $\pi_N$, as:
\begin{equation}
  \label{eq:TueMay14113228CEST2013}
  \pi_N={\lambda_N}/(\lambda_1+\lambda_N)\,.
\end{equation}
This ratio represents the percentage of $N$-photon bundles, that can
now be contrasted with $g^{(N)}$, as shown in Fig.~\ref{fig:3}(g) for
$N=2$. Here we find the remarkable result that $g^{(2)}$, often
described as the probability for two-photon emission, is in fact
anticorrelated with $\pi_2$, the actual such probability: when
$g^{(2)}$ reaches its maximum, $\pi_2$ is starting to grow and when
$\pi_2$ is maximum, $g^{(2)}$ is locally minimum, although still
larger than 1. 



We characterize the efficiency of $N$-photon emission by plotting the
purity and emission together, in Fig.~\ref{fig:4} for $\pi_2$ and
$\pi_3$, in the case $\gamma_\sigma/g=0.01$. Since $N$-photon emission
is a $(N+1)$th order process, it is more easily overcome by
dissipation as $N$ increases. However, almost pure two-photon and
three-photon emission is already feasible with state of the art cavity
QED systems: $\approx 85$\% of two-photon emission can be obtained in
semiconductor samples ($\gamma_a/g \approx 0.5$, $g \approx 12$
GHz)~\cite{ota11a,laucht09a} with a rate over $10^7$ counts per second (cps),
while circuit QED systems ($\gamma_a /g \approx 0.01$, $g \approx 50$
MHz)~\cite{nissen13a} can even reach $\approx 90$\% of three-photon
emission with a rate of $10^3$~cps.

Various regimes of $N$-photon emission can be characterized by
studying the statistics of the bundles when considered as single
entities. To do so we introduce the generalized correlation functions
$g^{(n)}_N$ as:
\begin{equation}
  \label{eq:mieoct9203726CEST2013}
  g^{(n)}_N(t_1,\ldots,t_n)=
  \frac
  {\langle\mathcal{T}_-\{\prod_{i=1}^na^{\dagger N}(t_i)\}\mathcal{T}_+\{\prod_{i=1}^na^{N}(t_{i})\}\rangle}
  {\prod_{i=1}^n\langle a^{\dagger N}a^N\rangle(t_i)}
\end{equation}
with $\mathcal{T}_\pm$ the time ordering operators. This upgrades the
concept of the $n$th order correlation function for isolated photons
to bundles of $N$ photons. The case $N=1$ recovers the definition of
the standard $g^{(n)}$~\cite{loudon_book00a}, but for $N\ge2$, the
normalization to the bundle density makes
Eq.~(\ref{eq:mieoct9203726CEST2013}) essentially different from the
standard correlation functions $g^{(n\times N)}$. Similarly to the
single-photon case, the two-bundle statistics
$g^{(2)}_N(\tau)=\frac{\langle a^{\dagger N}(0)a^{\dagger
    N}(\tau)a^N(\tau)a^N(0)\rangle}{\langle(a^{\dagger
    N}a^N)(0)\rangle\langle(a^{\dagger N} a^N)(\tau)\rangle}$ is the
most important one. The validity of this definition for $g^{(2)}_2$ is
confirmed in Figs.~\ref{fig:3}(h--i), where it is plotted (smooth
curve) along with direct coincidences between clicks from the Monte
Carlo simulation (data). Such $g_2^{(2)}$ correlations can be measured
thanks to recent developments in two-photon
detection~\cite{boitier09a}.  For the computation from the Monte Carlo
clicks, all events are considered as single photons for the standard
$g^{(2)}$ calculation (red curve in Fig.~\ref{fig:3}(h--i)), and only
two-photon events are considered as the basic unit of emission for
$g^{(2)}_2$ (blue curve). Except in the small jitter window of width
$1/\gamma_a$, photon pairs exhibit antibunching for long-lived QE
while they are Poisson distributed for short-lived QE. In the latter
case, one can also check that $g^{(3)}_2(\tau_1,\tau_2) = 1$ except
from the aforementioned jitter window.  The emitter therefore behaves
respectively as a two-photon gun, and---according to
Glauber~\cite{glauber63b}---as a laser (in the sense of a coherent
source), but at the two-photon level. At the single-photon level, the
standard $g^{(2)}(\tau)$ fails to capture this fundamental dynamics of
emission. The same behaviour holds for higher~$N$.

In conclusion, we have shown how to exploit cQED resonances in a
dissipative context to realize non-classical quantum sources that emit
most of the light in bundles of $N$ photons.  We introduced new
quantities to characterize such kind of systems, namely, the
percentage $\pi_N$ of the total emission that comes as $N$-photon
bundles, and generalized correlation functions $g^{(n)}_N$ that
overtake the standard definition in the regime of $N$-photon
emission. This allowed us to propose a whole class of versatile
$N$-photon emitters, ranging from quantum guns to lasers.

Work supported by the Spanish MINECO under contract MAT2011-22997 and by CAM under contract S2009/ESP-1503. C.S-M.~acknowledge a FPI grant and F.P.L.~a RyC contract.


\bibliography{Sci,books}

\end{document}